\documentclass{article}
\usepackage[T1]{fontenc}
\usepackage{iclr2027_conference,times}
\usepackage{amsmath,amssymb,booktabs,array,graphicx}
\usepackage{hyperref,url}
\usepackage{tikz}
\usetikzlibrary{arrows.meta,positioning,calc}
\hypersetup{hidelinks,pdftitle={From Verification Failures to Reusable Guidance for Coding Agents},pdfauthor={Yuqing Zhai; Xiaohong Chen; Lingming Zhang; Sriram Vishwanath; Grigore Rosu}}
\title{From Verification Failures to Reusable\\Guidance for Coding Agents}
\author{\textbf{Yuqing Zhai\textsuperscript{1} \quad Xiaohong Chen\textsuperscript{1}}\\
\textbf{Lingming Zhang\textsuperscript{2}\thanks{These authors took part in this work in collaboration with Intent Computing Inc., outside their university duties and without university funds, facilities, students or contracts.} \quad Sriram Vishwanath\textsuperscript{1,3}\footnotemark[1] \quad Grigore Rosu\textsuperscript{1,2}\footnotemark[1]}\\
\textnormal{\parbox[t]{\dimexpr\textwidth-2\tabcolsep\relax}{\raggedright \mbox{\textsuperscript{1}Intent Computing, Inc.} \quad \mbox{\textsuperscript{2}University of Illinois Urbana-Champaign} \quad \mbox{\textsuperscript{3}Georgia Institute of Technology}}}\\
\textnormal{\parbox[t]{\dimexpr\textwidth-2\tabcolsep\relax}{\raggedright \mbox{\href{mailto:yuqing@intentcomputing.org}{\nolinkurl{yuqing@intentcomputing.org}}} \quad \mbox{\href{mailto:xiaohong@intentcomputing.org}{\nolinkurl{xiaohong@intentcomputing.org}}} \quad \mbox{\href{mailto:lingming@illinois.edu}{\nolinkurl{lingming@illinois.edu}}} \quad \mbox{\href{mailto:sriram@ece.gatech.edu}{\nolinkurl{sriram@ece.gatech.edu}}} \quad \mbox{\href{mailto:grosu@illinois.edu}{\nolinkurl{grosu@illinois.edu}}}}}}
\newcommand{\K}{\mathbb{K}}

\newcommand{\code}[1]{\texttt{#1}}
\iclrfinalcopy
\begin{document}
\maketitle
\lhead{Preprint}

\begin{abstract}
Coding agents need to establish that a program satisfies a specification and that the specification captures the requested behavior. We study how expert diagnosis of verification failures can become reusable guidance for this work. Our approach combines executable language definitions in $\K$ with a kit of procedures for constructing specifications, repairing proofs, and auditing their adequacy. A human-guided development campaign on HumanEval, a benchmark of 164 Python programming tasks, achieves a 164/164 success rate with the semantics and the kit, measured by final AI audit \textsc{Pass} verdicts after two targeted repairs. To examine whether auditing detects problems that successful proofs leave unresolved, we construct 12 author-reviewed pairs of clean and defective packages. Every package passes its $\K$ proofs, and completed audits identify all defects and accept all clean packages. We then use KleverBench to test specification and proof construction for 31 programs with changed operator meanings. Comparisons with complete acceptance rules and equally long generic advice yield mixed results across two model and budget settings, motivating further work on selecting useful guidance within resource limits. Human-reviewed Optimism proofs establish expected pause reverts for six operations within declared input bounds under London semantics with unbounded gas. We report progress, difficulties, and lessons toward agents that deliver programs with checkable correctness arguments.
\end{abstract}

\section{Introduction}
\label{sec:intro}

Our long-term goal is to help coding agents deliver programs together with checkable correctness arguments. Stronger tests continue to expose errors in generated programs \citep{liu2023evalplus}, and reasoning about execution remains difficult for language models \citep{ni2024next}. Formal verification checks whether a stated property follows from explicit execution rules. The agent must also formulate a property that covers the requested behavior. A proof about a summation loop restricted to input zero, for example, leaves positive inputs unexamined. We ask how experience diagnosing such failures can be retained to guide later agents.

We use the $\K$ semantic framework to define how language instructions change the execution state \citep{rosu2010k}. The same definition supports concrete execution and symbolic proofs about inputs satisfying a stated condition \citep{stefanescu2016verifiers}. This gives agents a common verification method across languages, with rules for assignments, loops, and exceptions.

DafnyBench evaluates generation of missing proof hints \citep{loughridge2024dafnybench}, and AutoVerus organizes proof generation, refinement, and repair \citep{yang2025autoverus}. Clover checks consistency among code, descriptions, and formal annotations \citep{sun2024clover}. Reflexion retains textual reflections on feedback for subsequent attempts \citep{shinn2023reflexion}. We investigate how expert-curated procedures connect language semantics, specifications, proof extensions, and adequacy checks.

We retain this experience in shared semantics and a verification kit. Experts diagnose failures and record revised language rules or procedures with applicability conditions and checking examples. An invariant mismatch, for example, produces guidance for comparing the reached state with the proposed invariant. The kit supplies these procedures for specification, proof construction, and validation at inference time, with model weights fixed.

HumanEval provides 164 tasks for synthesizing Python functions from descriptions \citep{chen2021codex}. Our human-guided development campaign extends them to implementations, specifications, and proofs, using $\K$ for program reasoning and Lean for residual mathematical obligations \citep{demoura2021lean4}. With the semantics and the kit, we achieve a \textbf{164/164 success rate}, measured by final AI audit \textsc{Pass} verdicts after two targeted repairs. A separate challenge tests auditing on 12 author-reviewed clean/defective pairs. Every package's $\K$ claims prove, so the challenge asks whether audits detect defects in the modeled behavior and proof assumptions. Completed audits detect 12/12 defects and accept 12/12 clean packages, with 95\% Wilson intervals of 75.8--100\%.

KleverBench asks agents to construct specifications that pass structural, input-coverage, and $\K$ proof checks under supplied language rules. Its ordinary, renamed, and meaning-swapped definitions follow PLSemanticsBench \citep{thimmaiah2026plsemantics}. A subtraction symbol can, for example, denote addition. Two frozen comparisons use all 31 original programs under swapped semantics. Agents receive complete acceptance rules, those rules with generic advice, or those rules with a fixed kit core. The advice blocks have equal token lengths under the same tokenizer. With one attempt per program and condition, rules, generic advice, and kit acceptance is respectively 15/31, 15/31, and 19/31 with GPT-5.6 Luna, and 17/31, 18/31, and 16/31 with cash-capped DeepSeek-V4.1-Flash. All paired kit-effect intervals include zero. These mixed results make procedure selection and the cost of using guidance concrete questions for the next experiments.

Smart-contract studies apply the procedures to EVM semantics derived from KEVM \citep{hildenbrandt2018kevm}. They expose specification requirements for storage aliasing and call-level rollback. Human-reviewed Optimism proofs establish expected pause reverts for six operations within declared input bounds under London semantics with unbounded gas. Exploratory review on SWE-bench Verified examines repository repairs without formal tools \citep{jimenez2024swebench,openai2024verified}, as reported in Appendix~\ref{app:swe}. Our contribution is an implemented workflow for retaining expert verification experience and studies of how agents use it. We report the progress, difficulties, and lessons that guide further work on specification validation and reusable procedures.

\section{Programs, semantics, and proofs}
\label{sec:background}

\subsection{The verification task}
Let $P$ be a program, $\mathcal S$ an operational semantics, and $\varphi$ and $\psi$ predicates on initial and final states. Our target is partial functional correctness,
\begin{equation}
\forall s,t.\quad \varphi(s)\land \langle P,s\rangle\longrightarrow_{\mathcal S}^{*}
\langle\mathrm{done},t\rangle\quad\Longrightarrow\quad\psi(s,t).
\label{eq:correctness}
\end{equation}
Formalization chooses the semantics and predicates. Verification establishes the implication under the declared assumptions. Validation assesses whether those definitions capture the intended language behavior and requested property. This division follows the usual role of specifications in program correctness \citep{hoare1969axiomatic}. Equation~\ref{eq:correctness} applies to terminating executions. A termination theorem is a separate obligation.

When an agent produces its own specification, proof success is useful only in combination with evidence about that specification. Restricting the input domain to an easy case can omit the requested behavior. Adding a rewrite rule that directly asserts the desired result can make the checker accept an unjustified argument. We therefore retain the source program, its translation, the specification, every proof extension, and the evidence used to justify those extensions. Tests and independent audits help examine correspondence with the intended task, as in other work on consistency between code and specifications \citep{sun2024clover}.

\subsection{Executable semantics and domain reasoning}
We use $\K$ to define program execution and reason symbolically about it \citep{rosu2010k,stefanescu2016verifiers}. A configuration organizes the computation, variable environment, memory, and other runtime components into cells. Rewrite rules describe transitions between configurations. A reachability claim states that executions from an initial symbolic pattern reach a desired final pattern. The verifier combines symbolic execution with constraints and guarded circular \mbox{reasoning}, which can summarize a loop after execution has made progress \citep{lin2023proofcertificates}. The same semantics supports concrete execution for testing.

Consider a loop that starts with $n=N\geq0$ and $s=0$, and repeatedly adds $n$ to $s$ and decrements $n$ until it reaches zero. A useful loop summary predicts final sum $S+M(M+1)/2$ from a state with $s=S$ and $n=M\geq0$. After one iteration, symbolic execution gives $S'=S+M$ and $M'=M-1$. Reusing the summary leaves the arithmetic obligation
\begin{equation}
(S+M)+\frac{(M-1)M}{2}=S+\frac{M(M+1)}{2}.
\end{equation}
The language rules explain the state transition. Integer arithmetic explains why the summary is preserved. Keeping these obligations separate makes a failed proof easier to diagnose.

The HumanEval workflow uses Z3 for suitable domain implications \citep{demoura2008z3}. Additional domain lemmas are exported by \code{klean} as Lean theorem statements. The agent constructs their proofs, which Lean checks \citep{demoura2021lean4}. For example, a divisor proof uses the fact that integer remainder by one is zero. The trust boundary includes the source translation, language semantics, $\K$ and its solver integration, the lemma translation, and Lean's kernel and permitted axioms. Lean checks the exported domain proofs. The $\K$ execution proof and the correspondence of the exported statements remain distinct parts of the argument.

\section{Retaining verification experience}
\label{sec:kit}

\begin{figure}[t]
\centering
\begin{tikzpicture}[
 font=\small, >=Latex,
 box/.style={draw=black!65,rounded corners=2pt,align=center,
   minimum height=0.85cm,inner sep=4pt},
 flow/.style={->,line width=0.6pt,draw=black!65,
   >={Latex[length=2mm,width=1.4mm]}},
 guidance/.style={flow,dashed},
 feedback/.style={flow,rounded corners=2pt}]
\node[box,fill=blue!5,text width=3.8cm] (experience) at (0.7,1.6)
  {Failed attempts\\expert and user feedback};
\node[box,fill=blue!5,text width=3.1cm] (semantics) at (6.9,1.6)
  {\mbox{Executable semantics}\\rules and\\\mbox{conformance tests}};
\node[box,fill=blue!5,text width=2.65cm] (kit) at (10.35,1.6)
  {Verification kit\\procedures and\\references};
\coordinate (curation) at (3.85,1.6);
\draw[line width=0.6pt,draw=black!65] (experience.east)--(curation);
\draw[flow] (curation)--(semantics.west);
\draw[flow,rounded corners=2pt] (curation)--(3.85,2.75)
  --(10.35,2.75)--(kit.north);
\fill[black!65] (curation) circle[radius=1pt];

\node[box,text width=2.65cm] (task) at (0,0)
  {Program and\\intended behavior};
\node[box,text width=2.65cm] (spec) at (3.45,0)
  {Specification\\and scope};
\node[box,text width=2.8cm] (proof) at (6.9,0)
  {\mbox{Proof construction}\\and checking};
\node[box,text width=2.65cm] (audit) at (10.35,0)
  {Validation\\and evidence};
\draw[flow] (task.east)--(spec.west);
\draw[flow] (spec.east)--(proof.west);
\draw[flow] (proof.east)--(audit.west);
\draw[guidance] (semantics.south)--(proof.north);
\draw[guidance] (kit.south)--(audit.north);
\draw[feedback] (audit.south)--(10.35,-.78)--(3.45,-.78)--(spec.south);
\draw[feedback] (6.9,-.78)--(proof.south);
\node[font=\footnotesize,anchor=north] at (6.9,-.83)
  {repair the responsible stage};
\end{tikzpicture}
\caption{Experience is retained in executable semantics and verification procedures. Agents construct specifications and proofs. Validation examines programs, assumptions, and evidence. Feedback returns to the responsible stage. HumanEval also exports residual domain lemmas to Lean.}
\label{fig:workflow}
\end{figure}
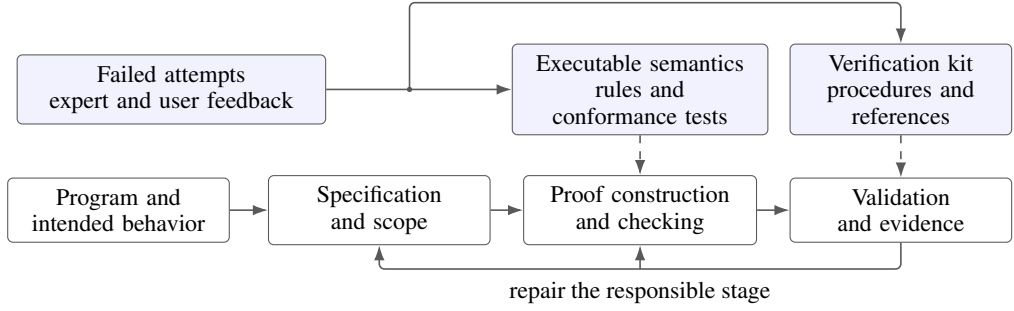

\subsection{From a failed attempt to a reusable procedure}
The initial HumanEval effort alternated proof attempts with expert diagnosis and repair. The first pass established language support and verification practices. The second pass organized generation, checking, and audit into a staged campaign. Experts inspected failures, diagnosed their causes, wrote procedures with applicability conditions, attached checking examples, and retained the changes in versioned resources. Semantic discrepancies produced revised language rules and tests that exposed the errors. Recurring proof failures produced procedures for subsequent agents. Figure~\ref{fig:workflow} separates this curation process from work on an individual task.

HumanEval's \code{below\_zero} function illustrates invariant diagnosis. Its result depends on whether any prefix of a sequence of balance changes has a negative sum. A proof must relate the processed prefix, remaining loop, and returned Boolean. Even a mathematically suitable invariant can fail to match the symbolic environment. The kit directs the agent to compare the residual computation, bindings, and framed state with the invariant, then preserve the mismatch and repaired claim as evidence. A proof-only replacement loop creates a separate obligation to establish equivalence with the source program.

The \code{split\_words} repair provides a traceable curation example. Its original predicate recognized four of CPython's 29 whitespace code points. The repair pins CPython and Unicode versions, implements all 29 members, and checks the table across every code point. The new kit entry requires a pinned library reference, complete finite-domain comparisons, boundary tests, and replays of affected consumers. The artifact links this instruction to the predicate change and independent checks in \code{kit/revisions/2026-09-24/}. Appendix~\ref{app:humaneval} records the repair evidence. The example records the instruction's provenance and checking procedure. Its effect on later tasks is unmeasured.

The current kit classifies proof extensions as mathematical definitions, derived lemmas, operational bridges, or declared trusted primitives. A definition introduces a mathematical value through equations. A derived lemma needs a derivation over its full guard. An operational bridge that replaces execution requires a machine-checked connection theorem over its full matched context, proved without that bridge. The theorem covers the state, bindings, and control affected by the replacement. This classification directs the next action. An unsupported result equation becomes a theorem to prove, and an overbroad bridge becomes a claim whose domain must be justified. A rejected false postcondition provides a further negative control.

\subsection{The current kit and plugin}
The current development snapshot packages nine skills and a shared reference library for coding agents. The live workflow selects a bundled semantics, writes the specification and its mathematical summaries, audits adequacy, constructs a proof, and requests a final proof audit. References on invariants, representation choices, and proof failures are loaded as needed. Stage outputs are claims, verification helpers, a scope document, replay commands, and audit reports. The released plugin contains eight skills, with the Rust command-line client, \code{kprover}, distributed separately to handle proof requests and evidence retrieval.

Validation checks whether proof extensions justify the claim, whether the claim covers the intended inputs and behavior, and whether the assumptions and checker evidence can be inspected and reproduced. A failed check returns work to the responsible construction stage with a specific obligation. Historical experiments retain their recorded acceptance criteria.

Contract-verification sessions exposed requirements for the tool interface. Compilation exceeded default time limits, diagnostics were missing, variable names collided with opcode tokens, and connections failed after a server had accepted a request. The current client records session identities, pins semantics sources, exposes task limits, and retrieves proof evidence. These changes address the observed execution failures. We evaluate historical runs with the tool versions they used.

\section{HumanEval with semantics and the kit}
\label{sec:humaneval}

HumanEval supplies 164 Python function tasks. In the second-pass protocol, the producer receives the task statement and its arm's resources, writes an implementation, and builds $\K$ specifications and proof artifacts. A shared frontend translates supported Python syntax into the semantics' term language. The arm with neither supplied resource receives the frontend and K tools and writes its own semantics. The Stage~2 auditor also receives the canonical implementation as behavioral evidence. Later stages classify proof extensions, export domain lemmas, construct Lean proofs when obligations are present, and audit the result. Table~\ref{tab:humaneval} reports \textsc{Pass}, \textsc{Concerns}, and \textsc{Fail} separately. The historical \textsc{Legit} metric combines the first two.

\begin{table}[ht]
\centering\small
\begin{tabular}{@{}lrrrrr@{}}
\toprule
Semantics / kit & Pass & Concerns & Fail & Pass rate & Legit rate\\
\midrule
Agent-authored / no kit & 23 & 41 & 100 & 14.0\% & 39.0\%\\
Supplied / no kit & 37 & 36 & 91 & 22.6\% & 44.5\%\\
Supplied / kit & 97 & 67 & 0 & 59.1\% & 100.0\%\\
\bottomrule
\end{tabular}
\caption{Historical HumanEval audits at the common Stage~2 endpoint, 164 tasks per arm. \textsc{Legit} combines \textsc{Pass} and \textsc{Concerns}. Resources, repairs, and attempt selection evolved during the campaign, preventing a controlled ablation. Final audit outcomes are reported separately.}
\label{tab:humaneval}
\end{table}

The revised artifact set has \textbf{164/164 final AI audit \textsc{Pass}} verdicts after two targeted repairs. Task 59 adds a Lean theorem establishing primality, divisibility, and maximality for its exact integer summary. Task 125 repairs the whitespace predicate against pinned CPython behavior. Fresh proof replays, discriminating negative controls, and independent AI repair audits validate both changes. The other 162 selected packages retain their historical evidence. Appendix~\ref{app:humaneval} preserves the original verdicts and the dated selection.

The historical producer is GPT-5.6 Sol at \code{xhigh}, and the repairs use a separately recorded toolchain. An author expert accepted both repairs under the review packet's stated assumptions and reported 30 minutes of review. HumanEval had already shaped the semantics and instructions, with repairs and continuations during the campaign. Generalization requires frozen evaluation on unseen tasks.

A separate study tests whether auditing detects gaps that successful proof replay leaves unresolved. An author expert reviewed 12 scoped clean/defective pairs before seeing the new AI judgments. The K claims in all 24 packages proved. Completed GPT-5.6 Sol audits identify all 12 seeded defects and accept all 12 clean packages. Each proportion has a 95\% Wilson interval of 75.8--100\%. One detected defect uses a local rule to bypass an incorrect body and fabricate its specified result. Human review takes approximately 60 self-reported minutes. Appendix~\ref{app:audit-calibration} gives the protocol, execution history, and results under the reviewed scopes.

The historical full-arm records contain 212 domain obligations across 72 tasks, while 92 tasks have none. Packages retain implementations, claims, source lemmas where present, and final audits. They expose recurring failures in invariant representation, input coverage, and helper-rule justification. Procedures for addressing these failures became the substance of the kit.

Shared semantics retains language behavior across tasks. An early remainder rule used the backend's Euclidean operation, giving $1$ where Python evaluates $7\bmod(-3)$ to $-2$. The repair and tests across operand signs retained that lesson. The kit guides use of these rules, source correspondence, and mathematical obligations. The semantics covers the Python subset exercised by the campaign. KleverBench then examines claim construction under changed programs and semantics.

\section{KleverBench under unfamiliar semantics}
\label{sec:klever}

\subsection{Tasks and acceptance checks}
KleverBench studies auto-formalization of supplied imperative programs. The historical comparison uses 31 programs under three semantics, yielding 93 program--semantics configurations. The ordinary semantics defines a small imperative language. The renamed variant replaces operators and keywords with unfamiliar glyphs while preserving their meaning. The swapped variant changes operator meanings behind familiar symbols. These transformations follow the semantic mutations of PLSemanticsBench \citep{thimmaiah2026plsemantics}, extending the question from predicting execution to constructing a specification that passes a verifier.

In this full-specification setting, the agent receives the program, semantics, and a fixed verification vocabulary. It must write reachability claims. Acceptance requires the main claim to contain the supplied program, characterize the final state in the prescribed form, and avoid forbidden proof rules or attributes. Concrete coverage points reject preconditions that exclude required inputs. The resulting claims must then reach \code{\#Top} under \code{kprove}. The mathematical vocabulary supplied by the benchmark is trusted for this experiment. This endpoint evaluates claims under that vocabulary and has a different scope from the HumanEval domain-lemma workflow.

\subsection{Frozen guidance with matched controls}
\label{sec:fixed-controls}
We test whether the kit's procedures improve acceptance beyond complete rules and generic advice on all 31 original programs under swapped semantics. Each program receives one attempt in each of three arms. Rules supplies the complete acceptance requirements. Generic adds advice about planning, checking, and revision. Kit instead adds a fixed historical core of procedures for scope, invariant derivation, circularity matching, residual inspection, and non-vacuity. The latter two additions each contain 2,145 tokens under the same named tokenizer. Every excerpt is supplied at the start. The arms share programs, semantics, mathematical vocabulary, checks, and time limits.

We repeat this design with GPT-5.6 Luna and DeepSeek-V4.1-Flash, using Codex 0.149.1, Harbor 0.22.0, and K 7.1.337. Each model has 93 original attempts in a randomized queue and a 30-minute agent limit. DeepSeek additionally has a \$0.10 request-admission allocation per trial. Task instructions are identical across models. Provider instructions and cash limits differ, so we interpret comparisons within model. Saved claims are graded at four workers on a quiet host using the historical proof limits. Table~\ref{tab:fixed-guidance} reports the original attempts. Appendix~\ref{app:fixed-controls} provides the frozen inputs, execution records, settings, and paired analysis.

\begin{table}[ht]
\centering\small
\begin{tabular}{@{}lrrrll@{}}
\toprule
Model & Rules & Generic & Kit & Kit--Rules & Kit--Generic\\
\midrule
Luna & 15/31 & 15/31 & 19/31 & $+12.9\ [-9.7,35.5]$ & $+12.9\ [-6.5,35.5]$\\
DeepSeek & 17/31 & 18/31 & 16/31 & $-3.2\ [-19.4,12.9]$ & $-6.5\ [-22.6,9.7]$\\
\bottomrule
\end{tabular}
\caption{Frozen guidance comparison on swapped semantics. Differences and pointwise 95\% paired bootstrap intervals are in percentage points, resampling 31 programs. Every original attempted cell remains in the denominator.}
\label{tab:fixed-guidance}
\end{table}

Luna's kit core accepts four more programs than either control, with eight paired wins and four losses against each. DeepSeek accepts one fewer than Rules and two fewer than Generic. All four intervals include zero. Within-model Holm-adjusted paired $p$-values are 0.775 for Luna and 1.0 for DeepSeek. The prespecified joint four-contrast adjustment gives 1.0 throughout. These controls make instruction content directly testable under complete rules. They also identify a practical resource question. DeepSeek reaches its local cash-admission limit in 14 Rules, 14 Generic, and 23 Kit trials. Saved candidates remain graded. Guidance selection under a fixed budget is therefore a concrete target for further study.

Luna's prospectively specified setup-recovery supplement gives 15/31, 16/31, and 20/31 accepted claims, with both kit-effect intervals including zero. Regrading its original and recovery candidates at one worker changes no verdicts. DeepSeek has no eligible setup recoveries. Appendix~\ref{app:fixed-controls} reports all first attempts, generation exceptions, resource records, and sensitivity analyses.

Luna's \code{add-loop} trajectories make the coverage issue concrete. Both controls prove claims restricted to $N\leq0$, which fail the independent input-domain check. The Luna kit candidate preserves the unrestricted domain and passes. On \code{count-even}, the kit instead narrows the domain to $N\leq0$, while both controls construct accepted loop claims. These examples show why input coverage needs its own check and why guidance must be evaluated through the claims that agents actually construct.

\subsection{Historical results and failure patterns}
The historical principal experiment uses Codex 0.149.1 with GPT-5.6 Luna and Harbor 0.22.0, K 7.1.337, and four attempts for each configuration and arm. Each arm contains 372 attempts. The agent time limit is 1,800 seconds. Grading allows 35 seconds per proof, increased to 100 seconds for two pipeline programs after checking reference-proof runtimes on the host. The kit version is pinned to the historical run. Our supplement independently aggregates per-configuration counts from the archived August 28 report. Trial trajectories are absent from the supplied snapshot, which supports count reconstruction and fresh runs from the archived setup.

\begin{table}[ht]
\centering\small
\begin{tabular}{@{}lrrrr@{}}
\toprule
& \multicolumn{2}{c}{Accepted attempts} & \multicolumn{2}{c}{Solved within four attempts}\\
Semantics & Bare & Kit & Bare & Kit\\
\midrule
Ordinary & 107/124 & 111/124 & 31/31 & 31/31\\
Renamed & 90/124 & 109/124 & 29/31 & 30/31\\
Swapped & 73/124 & 92/124 & 24/31 & 30/31\\
\midrule
All & 270/372 & 312/372 & 84/93 & 91/93\\
& 72.6\% & 83.9\% & 90.3\% & 97.8\%\\
\bottomrule
\end{tabular}
\caption{KleverBench's historical comparison. Attempt acceptance includes structural, coverage, and proof checks. Each base program appears under all three semantics. Counts retain failed attempts and timeouts.}
\label{tab:klever}
\end{table}

Table~\ref{tab:klever} shows 42 additional accepted attempts with the kit, an increase of 11.3 percentage points. A paired bootstrap over 31 base programs gives a descriptive 95\% interval of 6.2--16.7 points. The increase is 3.2 points under ordinary semantics and 15.3 points under each mutated semantics. On the original tiers 4 and 5, acceptance rises from 43/108 to 73/108, or 39.8\% to 67.6\%. The 27.8-point difference has an interval of 16.7--38.0 points across nine base programs. These tiers include nested loops and pipelines. Tiers 1 through 3 improve from 227/264 to 239/264. Appendix Figure~\ref{fig:klever} shows the tier proportions. The intervals group semantic variants by program.

Forbidden additions to the claims module appear in 28 bare attempts and zero kit attempts. The compatible archived bare prompt already prohibits rules and requires imports and claims only, although it omits some forbidden attributes. The reduction is consistent with better instruction adherence, clearer emphasis, or procedural guidance. The comparison cannot separate these explanations. Under swapped semantics, attempts that exclude required inputs fall from 31 to 26. For example, \code{sum-up} uses a subtraction symbol whose supplied meaning is addition. Some attempts restrict the input so the loop never executes. Concrete coverage checks reject that omission even when the restricted claim would prove.

A later extension provides nine translated programs under the same semantic variants. Across 108 attempts per arm, acceptance rises from 72 to 80. Ordinary semantics improves from 31/36 to 35/36 and renamed semantics from 27/36 to 31/36. Swapped semantics remains at 14/36 in both arms. This extension identifies unfamiliar operator meanings as a persistent difficulty. Appendix~\ref{app:klever} also retains the mixed archived DSV4-flash comparison under its different harness.

The principal report records 13,964 bare-agent turns and 6,764 kit-agent turns. Generation concurrency differs at four versus six, while grading uses four workers for both arms. Four cancelled bare attempts were replaced. Missing trajectories prevent replay of these attempts. Load and instruction compliance could explain part of the gain. Appendix~\ref{app:klever} reports the failure distribution and sensitivity calculations. The complete-rules and generic-advice arms in Section~\ref{sec:fixed-controls} directly address the instruction-content question raised by these records.

\section{Applying the kit to smart contracts}
\label{sec:contracts}

\subsection{Targets and completion}
The smart-contract study applies the kit to existing code using EVM semantics derived from KEVM \citep{hildenbrandt2018kevm}. Four historical examples supply the targets: HKG, DSToken, DSValue, and a storage-variable example. Fourteen final packages cover 15 function targets because one DSValue package treats two functions. The historical summary records selected initial outcomes and a human-requested follow-up on incomplete DSToken \code{transfer}. Newly recovered code-to-spec records match 12 final specifications byte for byte and expose repeated HKG runs. The final DSValue and transfer specifications differ from their recovered initial candidates. These counts describe retained packages. Selection rationale and unarchived attempts remain unknown.

Selected initial runs report 14/15 complete targets, 39 checked target claims, and one checked auxiliary claim. The final inventory contains 44 target and six auxiliary claims with proofs and passing final audits. Recovered transfer claims retain the final labels, but 11 of 12 statements change; DSValue's fork, gas, and calldata scope also changes. Appendix Table~\ref{tab:contracts} therefore uses the final inventory as a retrospective denominator. HKG's \code{totalSupply()} specifies rejection because the runtime lacks that getter. Initial token records yield an estimated AI cost of \$171.95, excluding follow-up and human effort. Appendix~\ref{app:contracts} separates recovered provenance, reference judgments, and unmeasured human work.

\subsection{What the transfer proof taught us}
DSToken \code{transfer} makes the specification problem concrete. The final package contains six raw-runtime claims and six message-call claims over the pinned runtime. Raw claims cover nonpayable rejection, the stopped condition, insufficient balance, overflow, and successful transfers with distinct or aliased storage slots. The message-call claims then establish the corresponding commit or rollback behavior. A raw exceptional execution can retain an earlier log entry in its state, while the enclosing failed call rolls back storage and logs. Both observations are consistent once the execution boundary is stated precisely.

These cases became guidance in the kit's EVM specification reference. Storage writes must be modeled in order, including cases where sender and receiver addresses alias. Uninitialized storage must read as zero. Event claims must follow the actual emitted memory slice and preserve any permitted initial log prefix. Failure behavior must follow the order of modifiers and runtime instructions. The \code{transfer} package fixes the Byzantium fork and disables gas accounting, so its claims concern functional state changes within that scope. These details show how practical verification experience can become reusable instructions for subsequent tasks.

\subsection{Pause behavior and the supported input domain}
The kit also guides specification and proof construction for six Optimism operations across the portal, standard bridge, ERC721 bridge, and cross-domain messenger. The agent receives source and implementation runtime bytecode. Its initial run proves six pause claims, with withdrawal proofs restricted to the empty array. Two feedback continuations incorporate saved proofs for lengths 1 through 10, with independent symbolic 600-byte elements. The final inventory has 16 operation claims and four separate claims about guardian pause and its propagation through dependency contracts. Backend records report proofs for all 20. Human review confirmed that the kit helped produce correct specifications and proofs for these scoped pause properties.

The operation claims establish \code{EVMC\_REVERT}, the expected pause-error payload, and preservation of target account state and initial logs. They execute direct implementation bytecode under London with unbounded gas. Dynamic byte arguments satisfy $\operatorname{length}(b)\leq2^{30}$, the inclusive bound of the bundled \code{\#bytes} ABI encoder. Bridge calls assume the authorized messenger and configured remote bridge. The pause-dependency claims establish the responses used by the operation proofs under specified storage and time conditions. These separate theorems account for the pause checks and their dependencies within the supported setup. Appendix~\ref{app:optimism} records the proof scope, human confirmation, and execution evidence.

\section{Related work}
\label{sec:related}

Reflexion retains verbal feedback, ExpeL extracts insights from experience, and Voyager stores executable skills for later use \citep{shinn2023reflexion,zhao2023expel,wang2023voyager}. Agent Workflow Memory induces reusable routines. Dynamic Cheatsheet and Agentic Context Engineering maintain evolving contextual guidance \citep{wang2024workflow,suzgun2025cheatsheet,zhang2025ace}. Our procedures are curated with expert involvement and stored as versioned instructions and references. They guide operational-semantics proofs and specify the evidence needed to justify mathematical summaries and proof extensions.

AutoSpec and SpecGen synthesize specifications with verification feedback \citep{wen2024autospec,ma2024specgen}. The nl2postcond study examines correspondence to natural-language intent and the ability of postconditions to reject incorrect behavior \citep{endres2023nl2postcond}. Lemur integrates language-model proposals with automated verification, and AlphaVerus uses translation and verifier-guided refinement \citep{wu2023lemur,aggarwal2024alphaverus}. VERINA separately evaluates code, specification, and proof generation \citep{ye2025verina}. These distinctions motivate our separation of checker acceptance from adequacy audits. We focus on reusable guidance for connecting executable semantics, mathematical summaries, and declared proof extensions.

Language models have been used for premise selection and proof construction in established formal environments. LeanDojo studies retrieval for Lean proofs \citep{yang2023leandojo}, and Baldur combines proof generation with repair from checker feedback \citep{first2023baldur}. DafnyBench evaluates generated verification annotations \citep{loughridge2024dafnybench}, while AutoVerus organizes generation, refinement, and debugging for Rust proofs \citep{yang2025autoverus}. Our work shares the use of checker feedback and staged proof construction. Its focus is the reusable experience needed to connect executable language definitions, generated specifications, proof extensions, and validation across different semantic settings.

Clover examines consistency among code, natural-language descriptions, and formal annotations \citep{sun2024clover}. This complements our use of audits and concrete checks to assess generated specifications. Work on execution reasoning shows the value of making program dynamics explicit \citep{ni2024next}. PLSemanticsBench demonstrates that changing the meaning of familiar symbols can disrupt model reasoning \citep{thimmaiah2026plsemantics}. KleverBench uses these mutations in a verification task requiring claims to pass adequacy checks and symbolic proof.

The underlying verification method comes from semantics-based program reasoning in $\K$ \citep{stefanescu2016verifiers}. KEVM demonstrates this approach for EVM execution \citep{hildenbrandt2018kevm}. Proof certificates can further reduce the trusted verifier implementation \citep{lin2023proofcertificates}. We build on these foundations through agent procedures, artifacts, and experiments. The present contribution concerns how agents use and validate formal reasoning in practice.

\section{Lessons and next steps}
\label{sec:discussion}

The studies show how a diagnosed verification failure can become a procedure with an applicability condition and an evidence requirement. An invariant mismatch calls for comparing symbolic states and framed bindings. A mathematical summary needs a theorem connecting it to the requested property. The HumanEval repairs make these requirements concrete through a summary-adequacy theorem and a checked library model. The DSToken case extends the same approach to raw execution and message-call rollback. The Optimism case applies these procedures to pause guards, symbolic ABI arguments, and dependency contracts. Each retained procedure states what an agent should inspect and what evidence should justify its next step.

The controlled experiments evaluate guidance beyond complete task rules and generic advice. Luna's kit count is higher and DeepSeek's is lower, with every paired interval spanning zero. These measurements give a reproducible baseline for improving procedure selection. The coverage traces and DeepSeek's more frequent kit-arm budget stops suggest two concrete experiments. Remove individual procedures to identify which instructions help preserve the input domain, and compare fixed excerpts with adaptive loading under several budgets. Selecting a new task family after freezing the resources would measure transfer with a clear exposure boundary. The present comparisons use a fixed historical core on existing programs whose prior exposure to kit editors is unknown.

The audit challenge supplies a complementary lesson about validation. All 24 packages prove their encoded claims. All completed audit judgments match the author-reviewed labels. The 12 selected pairs distinguish encoded proof success from correspondence to the intended behavior. Natural defects and additional human reviewers would extend the test to broader judgments. Repeating the historical audit protocol on the same labeled packages would separately assess the earlier campaign. Machine-checked translation and connection theorems would further reduce the judgments needed to relate source programs, mathematical summaries, and checked claims.

\label{maintext:end}
\subsection*{Reproducibility statement}
The supplement includes frozen instructions, generation records, candidate claims, grading logs, and code reconstructing the controlled comparison and audit results. It also retains historical HumanEval records, KleverBench reports and sources, SWE-bench outcomes, contract proofs, and kit snapshots. Offline scripts check identities, hashes, and aggregates without model access. Appendix~\ref{app:protocols} records versions, endpoints, and replay requirements. Fresh runs require the corresponding tools or services.

\subsection*{AI use statement}
Generative AI assisted experiment design and construction of programs, specifications, proof attempts, seeded-defect packages, and audit reports. It also assisted source inspection, result aggregation and interpretation, failure analysis, literature selection, artifact preparation, and manuscript drafting and revision. The artifact provides executable checks of aggregate results. AI audit verdicts are distinguished from formal checker judgments. The authors are responsible for the final scientific claims and submission.

\bibliography{references}
\bibliographystyle{iclr2027_conference}
\clearpage
\appendix
\section{Experimental protocols and evidence}
\label{app:protocols}

\subsection{HumanEval development and second-pass records}
\label{app:humaneval}
The first-pass semantics contains 23 modules, 2,121 lines, and 695 rewrite-rule declarations, with 39 focused semantics-test directories. The packaged second-pass semantics contains 23 modules, 2,365 lines, and 764 rule declarations. These counts describe different revisions. They should be associated with their respective tests and generated artifacts when reproducing a task. The supplement includes the historical frontend, semantics, instructions, and task files used by the packaged study.

The second-pass metadata records GPT-5.6 Sol at \code{xhigh} effort, Codex CLI 0.144.6, K and pyk 7.1.293, and Lean 4.22.0. Default construction budgets for K and Lean are 3,600 seconds initially and 7,200 seconds cumulatively. Classification has a default limit of 1,200 seconds. Selected task records include overrides, repairs, and rescue continuations, so realized compute varies by task. Producer, audit, and continuation costs belong to their respective task stages.

The historical \textsc{Legit} criterion accepts \textsc{Pass} and \textsc{Concerns}. Table~\ref{tab:humaneval} separates these verdicts at Stage~2 for every arm. The selected reviews give 23/41/100 \textsc{Pass}/\textsc{Concerns}/\textsc{Fail} for agent-authored semantics, 37/36/91 for supplied semantics, and 97/67/0 with the kit. Historical final Stage~6 audits contain 162 \textsc{Pass} and two \textsc{Concerns}. Formal checks establish claims under their definitions and assumptions. Audit verdicts also assess correspondence to intent. A benchmark-wide count of fully machine-checked intent adequacy is unavailable.

The two historical concerns identify gaps in tasks with empty domain-obligation lists. For HumanEval/59, \code{largest\_prime\_factor}, the original K proof connects execution to \code{lpfFrom}. Its largest-prime-factor property was still unproved. For HumanEval/125, \code{split\_words}, the original predicate recognizes four whitespace codes and omits several CPython whitespace characters. Original semantics, classifications, and reviews are preserved. The September 24 selection combines two repaired candidates with 162 unchanged candidates, yielding 164 final AI audit \textsc{Pass} verdicts. The aggregate checker verifies their source and evidence hashes.

Task 59 retains its original semantics version and adds a Lean adequacy proof. For every integer $n\geq2$, any interpretation of the three guarded summary equations returns a value at least two that is prime, divides $n$, and bounds every prime divisor of $n$. A well-founded argument proves the natural-number recurrence. An integer binding proves casts and correspondence for K's truncating division and normalized remainder. An explicit interpretation establishes that the equation premises are satisfiable. Independent source-only builds and axiom inspection find only \code{propext} and \code{Quot.sound}. Fresh K replays close the loop and entry claims. False targets and changed initial factors fail. The historical equations remain classified as definitions with an empty domain-obligation list. The new theorem discharges a separate adequacy obligation. Source transcription from K to Lean is independently audited and remains a trust assumption. K execution and Lean number theory are checked separately. The program claim establishes partial correctness.

Task 125 uses a versioned predicate matching all 29 whitespace members in CPython 3.12.3 with Unicode 15.0.0. Independent checks cover all 1,114,112 code points, including surrogates. Four original symbolic claims and 96 concrete K cases pass. False-postcondition and body-mutation controls produce meaningful failures. The source frontend accepts ASCII literals, while function inputs use a modeled integer-sequence character domain. Fresh inventory, export, and Lean preflight checks retain zero domain obligations and therefore have no Lean target theorem. Direct split/strip consumers are 19, 67, 91, 101, 125, and 143. Sensitivity replays pass for the other five tasks, which retain their selected historical semantics. Only task 125 adopts the new predicate.

The repairs use K 7.1.337 and Lean 4.24.0. For each task, one AI agent independent of its producer reconstructs the evidence and reviews both Stage~2 and final repair obligations. The agents see the original concern and producer evidence. These targeted audits leave the historical Docker audit pipeline and its records intact. Task 59's number-theory producer uses a separately recorded Claude session. Orchestration and auditing use the current coding-agent sessions. The artifact retains commands, durations, source hashes, and path-redaction records. Available producer usage covers part of the work. Total repair compute and other human effort remain unmeasured.

In a separate packet review, an author expert accepted both repairs under their stated assumptions and reported 30 minutes in total. The packet links requirements, statements, source correspondence, trust boundaries, and mechanical records. It excludes AI audit conclusions as review evidence. Prior discussions were visible, and the response does not specify a full line-by-line proof audit. The study below measures current auditing on other tasks without resampling the complete benchmark.

The supplementary checker recomputes the outcome counts and the residual-obligation distribution from the included records. It checks that all 164 task identifiers are present for every arm, preserves the distinction between an empty obligation list and a nonempty one, and links the task material to the recorded results. Its aggregate validation does not invoke the historical model or replay every K and Lean proof. The package's reproduction instructions distinguish those additional operations and their dependencies.

\begin{table}[ht]
\centering\small
\begin{tabular}{@{}>{\raggedright\arraybackslash}p{.15\linewidth}>{\raggedright\arraybackslash}p{.38\linewidth}>{\raggedright\arraybackslash}p{.39\linewidth}@{}}
\toprule
Audit & Inputs and mechanical evidence & Judgment and independence boundary\\
\midrule
Stage 2 & Candidate code, K artifacts, task statement, canonical implementation as behavioral evidence, and available checker results & Fresh model session examines semantics, specifications, and extensions. Canonical behavior is visible, so this is informed review.\\
\addlinespace
Stage 6 & Prior audit and classification, frozen target/source identities, inventories, obligation correspondence, clean Lean build, and axiom checks in proof mode & Fresh session with the recorded producer model family. Treatment condition and prior review are visible. Adequacy, classification, and bridge judgments remain model decisions.\\
\bottomrule
\end{tabular}
\caption{Historical audit protocol. Fresh sessions and read-only candidates separate audit from construction. The same model family is used and the treatment condition is visible. Empty-obligation tasks receive classification-only review.}
\label{tab:audit-boundaries}
\end{table}

Stage~6 records GPT-5.6 Sol at \code{xhigh}, starts a fresh session, and instructs it to distrust earlier passing verdicts. The supplied reviews record its judgments. Missing raw sessions and execution directories limit retrospective checks of protocol compliance. The targeted repairs' false-postcondition and body-mutation controls test checker behavior. The following prospective study separately tests an auditor's judgments.

\subsubsection{Author-reviewed seeded-defect audit}
\label{app:audit-calibration}
Before primary auditing, we froze 12 accessible HumanEval tasks, their scoped clean packages, paired seeded variants, neutral audit instructions, and analysis criteria. An author expert inspected the public requirements, scopes, exact changes, witnesses, and replay evidence. The reviewer accepted all proposed clean scopes and defects, saw only the review packet, and reported about five minutes per pair. The reviewer knew the proposed labels and saw no new AI judgments. This is one author-expert assessment. All 24 packages rebuilt and their claims reached \code{\#Top}, including every defective package. Concrete counterexamples and checked wrong-body or helper witnesses establish the seeded gaps under the recorded scope.

\begin{table}[ht]
\centering\small
\begin{tabular}{@{}lrrrr@{}}
\toprule
& & \multicolumn{2}{c}{Defects detected} & \\
Defect category & HumanEval tasks & Original & After retries & Clean accepted\\
\midrule
Weak postcondition & 41, 53, 138 & 3/3 & 3/3 & 3/3\\
Unsupported bridge & 54, 57, 120 & 2/3 & 3/3 & 3/3\\
Source/proof mismatch & 8, 48, 42 & 3/3 & 3/3 & 3/3\\
Incorrect semantics & 23, 49, 101 & 2/3 & 3/3 & 3/3\\
\midrule
Total & 12 pairs & 10/12 & 12/12 & 12/12\\
\bottomrule
\end{tabular}
\caption{The 24 original attempts include two capacity failures without reports. One additional attempt for each gives 24 completed judgments from 26 attempts. Every returned judgment matches the author-expert label. Clean acceptance is unchanged. All 24 packages passed their 28 K claims.}
\label{tab:audit-calibration}
\end{table}

Audits use GPT-5.6 Sol at \code{high}, Codex 0.155.0-alpha.16.3, K 7.1.337, and Python 3.12.3. Each fresh filesystem exposes one read-only candidate and local tools. Prior reviews, paired copies, seed labels, and producer narratives are absent. The same neutral prompt asks for source correspondence, scope coverage, semantic fidelity, and justification of local rules. No kit treatment is varied. Each session has a 20-minute cap and each K command a 180-second cap. At most two audits run concurrently with low scheduling priority and a shared lock serializing K commands. Their two-CPU affinity is nonexclusive and overlaps KleverBench generation. The frozen protocol assigns one attempt per candidate and counts nonresponses as unsuccessful. The artifact retains session logs, tool traces, recorded CLI model settings, and hashes.

The original 24 attempts identify 10/12 defects and accept 12/12 clean packages, with respective 95\% Wilson intervals [55.2, 95.3] and [75.8, 100]. The bridge variant of task 120 and semantics variant of task 23 return provider-capacity errors without reports. A post-primary amendment, committed before additional calls, permits one retry for each with unchanged candidates, prompts, models, and limits. Both sequential retries identify the defect. The completed-judgment view therefore gives 12/12 detection and 12/12 clean acceptance, each with interval [75.8, 100], across 26 total attempts. The original outcome remains primary, and no returned judgment is retried or replaced. Each metric has one observation per source task. An unblinded AI revision assistant judges every defect report to localize the seeded issue under the frozen rubric. This secondary assessment is separate from human ground truth and the mechanically counted auditor verdicts.

Human review totals approximately 60 self-reported minutes. Completed-turn records report 18,623,446 input tokens, including 17,250,304 cached tokens, and 179,709 output tokens. These CLI records provide partial usage accounting. The sample is small, selected for inspectable scope, and uses artificial defects. The protocol differs from historical Stage~6. The study neither calibrates all 164 selected artifacts nor replaces human review of the two repairs.

\subsection{KleverBench experimental versions}
\label{app:klever}

\subsubsection{Frozen guidance controls}
\label{app:fixed-controls}
The two studies use all 31 original \code{imp-swap} programs. Each freezes 93 task instructions, configurations, queue entries, source hashes, and the analysis protocol before generation. Task-instruction bytes match across models. The harness is \code{3ae2e43}, the kit source is \code{fe49fb5}, and the image contains K 7.1.337. Codex 0.149.1 runs through Harbor 0.22.0 at \code{high} effort with web search disabled. Sampling seed and temperature are unspecified by this adapter. The randomized queue uses seed 20260924. Excluded setup pilots use ordinary semantics. The provider catalog identifies the \code{deepseek-flash} alias as DeepSeek-V4.1-Flash. Provider instructions and Responses transport differ from Luna's.

Rules explicitly prohibits all seven rejected attributes and added definitions, requires the supplied program and symbolic state, and preserves the full terminating input domain. Generic advice concerns planning, evidence, revision, and completion. Kit advice comprises verbatim sections from five historical files, with retained source hashes and character ranges. The two advice blocks are each 2,145 tokens under \code{tiktoken} 0.14.0's \code{o200k\_base} encoder. Equality under the provider's internal tokenizer is unestablished. Shared benchmark rules take precedence when a general kit instruction permits an excluded action.

Each container has a four-CPU quota and an 8 GiB memory limit, and each agent has 1,800 seconds. Generation initially permits six concurrent trials. DeepSeek later uses five fixed credential slots, with at most one active trial per slot. Memory monitoring and admission guards prevent new launches under resource pressure. The artifact separates native agent time from scheduling delays. DeepSeek's remaining-only continuations follow diagnosed runtime and controller interruptions. Each amendment precedes its new calls and all grading, preserving every earlier first attempt, candidate, and charge. The historical proof cap is 35 seconds, with 100-second exceptions for two pipeline programs. Grading begins after generation and declared overlapping experiments finish. All saved candidates are graded at four workers. Luna additionally has a one-worker sensitivity, in fixed order with a shared compilation cache.

DeepSeek's proxy reserves an upper charge before each request against a \$0.10 allocation per cell. It limits outputs to 16,384 tokens and requests to 240,000 bytes, uses peak uncached input rates, and retains reservations when terminal usage is missing. The 93 cells allocate at most \$9.30; setup has its own allowance. A local admission stop retains the available candidate for grading and stays in the denominator. Equal allocations can produce different consumed compute across arms. We report conservative charges, usage, and admission stops; invoice costs are unavailable.

The primary contrast is Kit minus Rules, and Kit minus Generic is secondary. Each difference has a pointwise percentile interval from 10,000 bootstrap resamples of the 31 programs using seed 20260924. Exact two-sided McNemar tests use paired discordances, with Holm correction over the two contrasts within each model. A prespecified joint secondary correction covers all four contrasts across models. Recovery supplements retain separate labels and correction families. The artifact retains the original protocols, analysis implementations, and dated amendments. The intervals describe variation across observed programs and exclude repeated-generation variance within a task.

\begin{table}[ht]
\centering\small
\begin{tabular}{@{}llrrrr@{}}
\toprule
Model / sample & Contrast & Kit-only & Control-only & Exact $p$ & Holm $p$\\
\midrule
Luna original & Kit--Rules & 8 & 4 & .3877 & .7754\\
Luna original & Kit--Generic & 8 & 4 & .3877 & .7754\\
Luna recovery & Kit--Rules & 8 & 3 & .2266 & .4531\\
Luna recovery & Kit--Generic & 7 & 3 & .3438 & .4531\\
DeepSeek original & Kit--Rules & 3 & 4 & 1.0000 & 1.0000\\
DeepSeek original & Kit--Generic & 2 & 4 & .6875 & 1.0000\\
\bottomrule
\end{tabular}
\caption{Paired comparisons on 31 programs. Original attempts are primary. Holm correction uses two contrasts within each model or recovery supplement. The joint four-contrast primary correction gives 1.0 throughout. DeepSeek has no eligible recoveries.}
\label{tab:fixed-guidance-tests}
\end{table}

Four original Luna trials fail during installation before any model call. A dated amendment, fixed before primary grades, permits at most one additional setup attempt for each eligible cell. The original 93 attempts remain primary. This supplement accepts 15 Rules, 16 Generic, and 20 Kit candidates, each out of 31. Kit--Rules is $+16.1$ points with interval $[-3.2,35.5]$, and Kit--Generic is $+12.9$ with interval $[-6.5,32.3]$. The original sample restricted to the 27 program blocks without setup exceptions gives 13, 14, and 18 acceptances. Its intervals also cross zero.

The recovered \code{rows-sum-nested} Generic trial leaves a candidate before its agent exits with status 143. Its final cleanup command matches the agent's own command line, indicating likely self-termination. The candidate passes structural checks and times out in both proof regrades. The prespecified exception filter excludes this block despite its likely model-triggered interruption, giving 15, 16, and 20 acceptances out of 30. Full recovery results retain this failure.

No original or selected recovery verdict changes when grading uses one worker. Candidate hashes and every structural check also remain unchanged. Among the 49 original candidates proving under both settings, median proof time changes from 3.3 to 2.3 seconds. This runtime comparison excludes censored timeouts and has fixed pass order. The result provides a reproducible concurrency sensitivity for the new candidates. Missing historical trajectories prevent the same check for the earlier 28-versus-four timeout difference.

Before grades, we specified the first alphabetical kit-only success and kit loss as exploratory examples. These are \code{add-loop} and \code{clamp}. The latter is an installation failure, and its recovery passes. We additionally inspect \code{count-even}, the first alphabetical loss with a model-generated kit candidate. The supplied traces link these examples to scope restrictions and loop claims. This additional illustration was selected after grades and supplies no procedure-level causal estimate.

DeepSeek completes all 93 first attempts. Rules, Generic, and Kit accept 17, 18, and 16 candidates. Kit--Rules is $-3.2$ points with interval $[-19.4,12.9]$, and Kit--Generic is $-6.5$ with interval $[-22.6,9.7]$. An installation interruption makes no model call but falls outside the frozen recovery classifier. It remains a failed original attempt. The declared exception filter removes four entire program blocks, giving 14, 17, and 16 acceptances out of 27. The corresponding differences are $+7.4\ [-7.4,22.2]$ and $-3.7\ [-18.5,11.1]$, with both Holm-adjusted $p=1.0$. One excluded trial follows a broad model-issued process-termination command, suggesting self-termination. The filter includes a possibly model-caused interruption; the primary analysis retains all original candidates.

Local cash-admission stops affect 14 Rules, 14 Generic, and 23 Kit trials. A refused reservation can occur below the nominal allocation because it covers a full possible response plus outstanding requests. Available ledgers account for an upper charge of \$4.06395. Including the missing-ledger allocation and excluded setup smoke gives a conservative bound of \$4.18075. Actual invoiced cost is unavailable. Generation takes 42.38 summed minutes across eight execution segments, excluding preparation gaps. Quiet grading takes 4.20 minutes. Minimum sampled available memory is 17.06~GiB during generation and 16.39~GiB during grading, with no sampled pressure alerts. The artifact retains per-arm timing, usage, admission records, and all operational amendments.

\subsubsection{Historical comparisons}
The primary report is dated August 28, 2026. Its first committed source snapshot, \code{3ae2e43}, retains 31 programs under each of three semantics. The report identifies kit revision \code{fe49fb5}, K 7.1.337, and container image \code{kleverbench-k:7.1.337-18e1a69628cf}. It records a 16-core Xeon Skylake host with 30 GB RAM, native amd64 execution, and Docker 29.2.1. The exact source revision at generation time, sampling seeds, temperature, and token budget are unrecorded. The included historical tree is the compatible report snapshot, with that provenance stated in its manifest.

Before generation, reference-proof timing motivated a grading limit of 35 seconds for most programs and 100 seconds for \code{pipeline} and \code{stats-pipeline}. All 93 reference configurations then proved, with a reported slowest reference of 54.6 seconds. Generation concurrency was four for bare and six for kit. Final grading used \code{--jobs 4} for both. Four cancelled bare trials were replaced after an interrupted batch. Twenty-seven bare agents and eleven kit agents reached their agent time limit. Their last saved claims were graded, and four kit claims from capped agents proved. Agent-cap counts overlap proof outcomes and describe a different event from final grading timeouts.

\begin{table}[h]
\centering\small
\begin{tabular}{@{}lrr@{}}
\toprule
Primary comparison outcome & Bare & Kit\\
\midrule
Accepted & 270 & 312\\
Rejected by pre-proof checks & 66 & 34\\
Proof timeout & 28 & 4\\
Proof error & 8 & 22\\
\midrule
All attempts & 372 & 372\\
\bottomrule
\end{tabular}
\caption{Disjoint reported outcomes. Individual pre-proof check failures can overlap within an attempt and should not be added to this table as additional outcomes.}
\label{tab:failures}
\end{table}

Across the 93 program--semantics configurations, the kit produces more successful attempts in 34, the same number in 53, and fewer in six. Both arms solve 83 configurations at least once. The bare arm alone solves one, the kit alone solves eight, and neither solves one. These are configuration-level comparisons reconstructed from per-problem counts. The original harder tiers contain nine base programs, each repeated across three semantics and four attempts. Later repository versions assign tiers using observed outcomes. We retain the original report's authored difficulty labels.

\begin{figure}[ht]
\centering
\includegraphics[width=\linewidth]{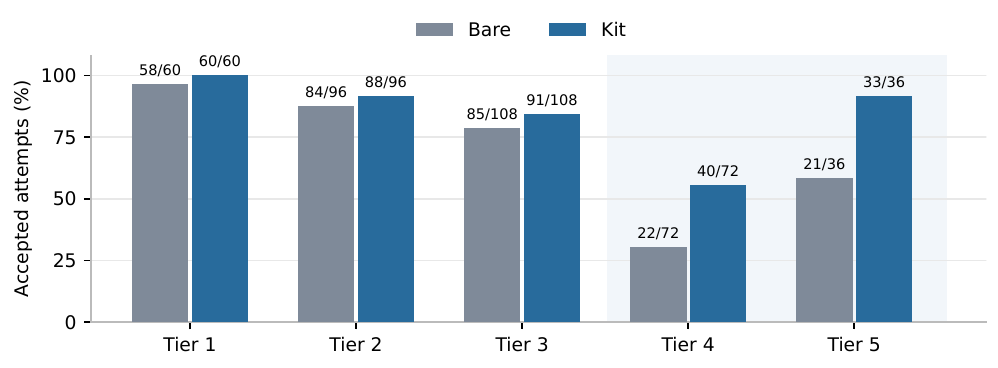}
\caption{Acceptance by the five original authored difficulty tiers in the principal comparison. Each tier combines all three semantics. Tiers 4 and 5 include nested loops and pipelines. The bars show observed proportions across 372 attempts per arm.}
\label{fig:klever}
\end{figure}

The analysis bootstraps paired base programs, keeping both arms and all three semantic variants together. For each cohort it draws 10,000 samples with replacement, resets the random seed to 20260924, and takes the 2.5th and 97.5th percentiles with linear interpolation. The overall 31-program difference is 11.3 percentage points with interval [6.2, 16.7]. The nine original harder-tier programs give 27.8 points with interval [16.7, 38.0]. The retained counts suffice despite missing trajectories. These intervals describe variation across the observed programs. Run-level uncertainty, exposure, and causal attribution require further evidence.

\begin{table}[ht]
\centering\small
\begin{tabular}{@{}>{\raggedright\arraybackslash}p{.40\linewidth}>{\raggedright\arraybackslash}p{.52\linewidth}@{}}
\toprule
Acceptance constraint & Compatible archived bare instructions\\
\midrule
Imports and claims only, no rule/syntax/configuration/context items & Explicitly stated. The agentic renderer incorporates these clauses.\\
Exact supplied program, symbolic initial state, no final existential weakening & Explicitly stated.\\
Coverage of required inputs & Weak preconditions requested, concrete coverage enforced by the checker.\\
Forbidden claim attributes & \code{trusted}, \code{simplification}, \code{macro}, \code{alias}, \code{anywhere}, \code{owise}, and \code{priority} are checked but not individually enumerated in that prompt.\\
\bottomrule
\end{tabular}
\caption{Prompt/checker correspondence, reconstructed from the compatible source snapshot. Historical rendered prompts and generation HEAD are unavailable, so this reconstruction cannot establish their exact content.}
\label{tab:prompt-contract}
\end{table}

Historical kit guidance permits simplification rules in general specification work. The benchmark imposes stricter requirements. The bare prompt already states the claims-only rule, so reduced violations could reflect adherence or emphasis. Table~\ref{tab:failures} attributes the 42 additional acceptances arithmetically to 32 fewer pre-proof rejections and 24 fewer timeouts, offset by 14 more proof errors. Crediting all 28 bare timeouts gives 298/372 versus 312/372. Crediting those and the 28 disjoint claims-only rejections gives 326/372 versus 312/372. These are hypothetical acceptance counts. Overlapping pre-proof subcategories prevent unique attribution. The frozen comparison in Section~\ref{sec:fixed-controls} supplies these two controls for a fixed historical excerpt.

\begin{table}[h]
\centering\small
\begin{tabular}{@{}lrr@{}}
\toprule
Nine-program extension & Bare & Kit\\
\midrule
Ordinary semantics & 31/36 & 35/36\\
Renamed semantics & 27/36 & 31/36\\
Swapped semantics & 14/36 & 14/36\\
\midrule
All & 72/108 & 80/108\\
\bottomrule
\end{tabular}
\caption{Separate September 4 report, after the language and prompt vocabulary had changed. These attempts are not pooled with the primary experiment.}
\label{tab:extension}
\end{table}

The later translated cohort contains common factors, cubes, a count-and-fill program, Euclidean gcd, an array inversion, an array access-and-sum program, digit reversal, trailing zeroes, and tribonacci. Its report identifies data revision \code{c738bd4}, the same kit revision, and a changed container image and prompt vocabulary. The current benchmark expands to 45 reference programs per semantics and 71 gold-post programs. The supplement includes the current sources and tests alongside the historical comparison. The presence of new programs in the current tree is distinct from measured outcomes on those programs.

The archived August~28 DSV4-flash study uses OpenCode 1.18.23 and one attempt per configuration, with concurrency eight in both arms. Acceptance is 67/93 bare and 70/93 kit. Swapped semantics falls from 26/31 to 22/31. The different harness limits comparison with Luna. Model capability ordering is unestablished, so this result leaves the weaker-model question open. The separate skill-use report concerns DSV4. Of its 93 trials, 86 open \code{using-kit}, 21 open \code{writing-spec}, and seven open \code{proving-spec}. Instruction loading in the principal Luna trials is unrecorded.

\subsection{Semantics transformations and adequacy checks}
The swapped semantics exchanges addition and subtraction, multiplication and division, and conjunction and disjunction at the object-language level. Comparison symbols, modulo, and indexing retain their meanings. Programs are transcribed along with the semantics so the reference behavior remains aligned. The renamed semantics substitutes unfamiliar glyphs for operators and keywords. Negative-literal signs and K's mathematical operators, such as \code{+Int}, are preserved. Tests check program and claim transcription, including helper-function bodies stored in the configuration.

For example, a transcribed \code{sum-up} program initializes \code{sum = 0} and \code{i = 1}, then executes \code{sum = sum - i} and \code{i = i - 1} while \code{i <= n}. Under the swapped meanings, this accumulates the ordinary triangular sum and increments \code{i}. For $N\geq0$, the final values are $\mathrm{sum}=N(N+1)/2$ and $i=N+1$. A claim restricted to $N<1$ avoids loop execution and fails to cover the input $N=1$. The coverage check rejects that omission. It checks the supplied finite set of inputs, which is useful evidence about domain coverage without establishing a weakest precondition.

The gold-post evaluation executes coverage inputs with the supplied semantics to obtain positive final states. It perturbs named outputs to generate candidate negatives and keeps those rejected by the hidden gold postcondition, up to 40 per problem. The model's precondition must accept coverage inputs, and its postcondition must accept positives and reject negatives. The symbolic claims must separately prove. Agents may supply verification helpers in this evaluation, so their validity is an additional assumption to inspect. The bare-agent report records 71 delivered modules, 60 passing the checks, and 41 proving within a 300-second grading limit.

The benchmark tests exercise forbidden proof rules and attributes, program identity, existential weakening, coverage-point handling, and legal case splits. Execution tests cover arrays, helper functions, and loop breaks. Break tests include guard-false and break exits and reject a specification that covers only the false guard. Concrete outcomes in the newer evaluation come from K execution, keeping the oracle aligned with the supplied semantics. The supplement's quick checks exercise data aggregation and selected structural tests. Replaying executable semantics and proofs requires the recorded K toolchain.

\subsection{Smart-contract scope and evidence}
\label{app:contracts}
A function target is complete when its retained candidate package has proofs for all its stated claims. A target claim describes requested behavior, while an auxiliary claim supports a proof. The final historical corpus contains 14 packages, 15 function targets, 44 target claims, and six auxiliary claims. The selected initial summary reports 13 complete packages and 14 targets. The follow-up completes transfer, while comparison with recovered initial sources also shows scope changes in DSValue. All retained runs use the kit. Estimating its incremental effect requires a matched no-kit comparison. Token-derived cost estimates concern the selected initial runs and exclude follow-up, human feedback, and historical reference development.

\begin{table}[h]
\centering\small
\begin{tabular}{@{}lrrrr@{}}
\toprule
& \multicolumn{2}{c}{Selected initial runs} & \multicolumn{2}{c}{After targeted feedback}\\
Example & Complete targets & Checked claims & Complete targets & Checked claims\\
\midrule
HKG & 6/6 & 15/15 & 6/6 & 15/15\\
DSToken & 5/6 & 18/28 & 6/6 & 28/28\\
DSValue & 2/2 & 5/5 & 2/2 & 5/5\\
Storage variable & 1/1 & 2/2 & 1/1 & 2/2\\
\midrule
Total & 14/15 & 40/50 & 15/15 & 50/50\\
\bottomrule
\end{tabular}
\caption{Selected historical smart-contract packages. Claim denominators use the final inventory. Recovered initial transfer and DSValue statements differ from their final versions, so these are not matched-claim completion rates. Follow-up completes one function target. Counts concern the scopes in the retained packages.}
\label{tab:contracts}
\end{table}

The original snapshot contains selected final packages and initial aggregate metrics. The newly recovered archive supplies 20 listed GPT-5.6 Sol code-to-spec runs across these four examples, including two HKG runs per function, plus 14 DeepSeek target-result rows and retained interrupted attempts. Twelve of the 14 historical final \code{spec.k} files match Sol candidates exactly: six HKG, five DSToken, and storage. The source ledger records every match. Initial and final DSValue claims retain five labels but change the schedule, gas model, and trailing-calldata domain. Initial and final transfer retain 12 labels, but 11 statements change. The recorded initial nonpayable proof pair therefore does not turn the reported 40/50 into completion of 40 identical final statements. The historical selection rationale and total off-record attempts remain unknown. Recovered prompts withhold reference specifications from constructors, while training exposure remains uncontrolled.

The recovered records separate backend proof, independent kit audit, and reference-judge verdict. Both DSValue constructors prove their candidate claims but receive reference \textsc{Fail} for narrower execution scopes. Sol's DSToken \code{transferFrom} receives proof, audit, and judge success, yet source comparison exposes its restriction to empty initial logs. A judge \textsc{Pass} therefore does not certify reference-domain coverage. The matching initial HKG \code{balanceOf} run has no completed independent audit before timeout, while the historical final package records a separate validation summary whose timing is unrecorded. The supplement retains unfavorable and missing results, HKG repetitions, and a separate alternative-prompt cohort rather than pooling them into a success rate. Newly recovered trajectories improve provenance without supplying a no-kit comparison.

\begin{table}[ht]
\centering\small
\begin{tabular}{@{}lrrr@{}}
\toprule
Selected initial activity & Input tokens & Cached input & Output tokens\\
\midrule
HKG & 157,550,964 & 154,774,528 & 422,205\\
DSToken & 149,267,850 & 146,764,928 & 427,152\\
DSValue & 12,624,914 & 12,379,136 & 55,618\\
Storage variable & 10,883,735 & 10,631,936 & 45,900\\
\midrule
Total & 330,327,463 & 324,550,528 & 950,875\\
\bottomrule
\end{tabular}
\caption{Recorded selected initial-run usage. Input includes cached input, and output includes reasoning tokens. Follow-up usage, human minutes, expertise, and the complete run census are unavailable. The source estimates \$171.95 for the measured activity. A cost comparison with manual verification is unavailable.}
\label{tab:contract-effort}
\end{table}

Recorded proof/verification medians in minutes are 37.7/63.0 for HKG, 16.8/46.8 for DSToken, 22.5/39.0 for DSValue, and 13.3/29.4 for storage. Proof medians include completed initial candidates, excluding the incomplete transfer. Verification medians include all measured targets. DSValue measures its two functions together. Total development time and human labor cannot be recovered from these medians.

The DSToken transfer proof fixes a 6,955-byte runtime. Raw-execution cases cover a nonzero call value, a stopped token, insufficient balance, overflow, and successful transfers to distinct or aliased slots. Call-level claims establish commit or rollback and the call-success flag. Scope files record ABI constraints, storage and log assumptions, and fork and gas settings. Transfer uses Byzantium with gas disabled. Other packages have their own schedules, and DSValue uses positive-infinite gas with cost tracking. The proofs cover the specified behaviors within these scopes.

Each package includes its specification, supporting claims, scope, final proof report, audit material, and a replay command. The submission's analysis checks package counts, labeled claims, recorded completion, and bytecode identities. It does not claim a new execution of the remote prover. Replay requires a compatible Prover service and semantics revision, as explained in the artifact. An audit report and an independently rerun proof are distinct forms of evidence, and the supplementary records identify which is provided.

\subsection{Optimism pause behavior and proof scope}
\label{app:optimism}
The Optimism case supplies Solidity sources and compiled implementation runtimes from revision \code{33fbe016} to GPT-5.6 Sol at \code{xhigh}, using Codex 0.153.4, KProver 0.1.1, kit \code{8a2f727}, and bundled EVM semantics \code{4f4c384}. The agent has a 12-hour limit and up to 100 proofs and 100 validations per Prover session. Runtime Verification's reference specifications are withheld from the constructor and supplied to a separate GPT-6 Astra judge at \code{max}. One initial run and two feedback continuations retain the same work. This continued effort has no matched no-kit comparison.

The author reports a human review and check confirming that the kit helped generate correct specifications and proofs for the scoped pause behavior. The supplement retains that confirmation in \code{human-review.json}. This confirmation concerns the supported input domain and execution assumptions documented below. The archived automated assessments and execution records remain available as provenance for the case history.

\begin{table}[ht]
\centering\small
\begin{tabular}{@{}llr@{}}
\toprule
Implementation & Operation & Claims\\
\midrule
OptimismPortal2 & \code{proveWithdrawalTransaction} & 11\\
OptimismPortal2 & \code{finalizeWithdrawalTransaction} & 1\\
L1StandardBridge & \code{finalizeBridgeETH} & 1\\
L1StandardBridge & \code{finalizeBridgeERC20} & 1\\
L1ERC721Bridge & \code{finalizeBridgeERC721} & 1\\
L1CrossDomainMessenger & \code{relayMessage} & 1\\
\midrule
Operation subtotal & Six initial and ten array-extension claims & 16\\
Pause dependencies & Guardian write and three pause getters & 4\\
\bottomrule
\end{tabular}
\caption{Retained Optimism claims with positive backend proof records. Eleven withdrawal claims cover lengths 0--10, with independent symbolic 600-byte elements for nonempty arrays. The original six claims are counted once. Human review confirms the specifications and proofs under the documented input and execution assumptions.}
\label{tab:optimism}
\end{table}

The portal claims return the custom error \code{OptimismPortal\_CallPaused()}. The standard-bridge targets return \code{StandardBridge: paused} through \code{Error(string)}. The ERC721 and messenger targets similarly return \code{L1ERC721Bridge: paused} and \code{CrossDomainMessenger: paused}. Every operation claim requires \code{EVMC\_REVERT} and preserves target account state and initial logs. The operation claims use direct implementation runtimes, the London schedule, empty call stacks, symbolic call depth $0\leq d<1023$, and positive-infinite gas with gas tracking enabled. Canonical ABI data and each claim's scalar range and account constraints remain assumptions. Bridge calls require the authorized messenger and configured remote bridge. Malformed calldata, finite-gas outcomes, and deployed proxies are outside this scope.

All 16 operation claims restrict their variable byte arguments to length at most $2^{30}$, inclusive. The pinned \code{abi.md} (line 456) imposes the condition \code{lengthBytes(BS) <=Int 1073741824} on dynamic byte encoding through an \code{ensures} clause. The supplement retains the full source and its hash. This is a restriction of the helper used for the proof inputs, not a Solidity contract bound or a limit on all raw EVM calldata. The reference documentation states a $2^{63}$ byte bound. The final judge identifies the larger byte domain as the sole remaining reference-coverage gap, while accepting the expected errors, authorization, array cases, and separately organized pause dependencies. That judgment supplies no theorem extending a claim beyond its written guard.

The original six claims have a retained combined proof, independent replay, and a wrong-status negative control. The first continuation completes only one of ten new array claims; its timeouts and incomplete attempts remain in the archive. The final continuation incorporates ten saved positive array-proof records supplied through feedback. Its array auditor validates their correspondence but does not independently replay them; a separate validation remained queued and was cancelled. The audit reports \code{SOUND-BUT-LIMITED}, with failed adequacy and replay-evidence gates. Four further positive proof records establish a guardian timestamp write and active-pause responses from SuperchainConfig, ETHLockbox, and SystemConfig under their storage, wiring, and timestamp assumptions. These four claims have no separate proof audit. They do not compose the guardian and all six operations into one execution theorem. The submission's offline checks inspect the retained records and claim identities; no new remote proof run is claimed.

\subsection{Exploratory SWE-bench review}
\label{app:swe}
\label{sec:swe}
We study kit-guided review on the 500-task SWE-bench Verified subset \citep{jimenez2024swebench,openai2024verified}. A second pass continues from a baseline patch and uses verification procedures to inspect assumptions and execution paths. Generation prompts prohibit tests and formal-tool calls. The benchmark harness evaluates both patches afterward. This study records behavioral-review cases for further investigation.

The campaign comprises 50 batches covering 500 unique tasks. Its manifests identify GPT-5.5 at \code{xhigh} effort and Codex CLI 0.140.0. The 470 tasks using kit \code{fef0123} give 380 baseline and 385 reviewed passes, with seven gains and two losses. The 30 tasks using \code{25a33f4} give 27 and 28 passes, with one gain. Different tasks were assigned to the two versions, preventing a version-effect estimate. A duplicate later batch is excluded. Its six shared passes explain the older count of 411 as 405 canonical plus six duplicate cases. It contributes no additional changed shared-pass patches.

\begin{table}[h]
\centering\small
\begin{tabular}{@{}lr@{}}
\toprule
Paired evaluator outcome & Tasks\\
\midrule
Both patches pass & 405\\
Both patches fail & 85\\
Baseline fails, reviewed patch passes & 8\\
Baseline passes, reviewed patch fails & 2\\
\midrule
Total & 500\\
\bottomrule
\end{tabular}
\caption{The stored evaluator reports give 407 baseline passes and 413 reviewed passes. The earlier tracker records 415 reviewed passes because it marks both regressions as solved. The supplement preserves and checks these two disagreements.}
\label{tab:swe}
\end{table}

Among 405 shared passes, review changes 86 patches. The case corpus labels 60 as candidate baseline issues and excludes 26. Severity judgments comprise nine high, 21 medium, and 30 low. Three candidates have archived added-test confirmations. Xarray 4094 concerns a length-one dimension across stack/unstack, scikit-learn 13496 concerns positional constructor arguments, and Sphinx 9367 concerns a singleton-tuple comma. Each baseline fails the added test and its reviewed patch passes. Nine other reports describe execution checks, while 48 rely on source inspection or hand reasoning. The ledger records the evidence tier and disposition for all 86 cases.

Fresh targeted execution covers all nine cases with historical execution narratives. Unchanged baseline, reviewed, and gold patches run on each task's base source commit in matched, pinned environments. Six cases confirm behavior supported by the issue or pre-patch API documentation. Django 13569 confirms SQL compilation structure only. Requests 2931 and SymPy 24066 reproduce private-helper or unsupported-input differences whose defect interpretation remains unresolved. All nine reviewed patches pass the focused checks. Gold misses some additional supported behaviors, so expected outcomes also require independent intent evidence. The artifact retains 27 selected executions and three initial settings-setup errors, followed by a complete matched rerun for that case. These selected tests cover the known narratives. The other 51 candidate issues retain their previous evidence. Full official suites were not rerun.

The tracker marks Sphinx 8056 and SymPy 12419 as resolved by both routes. Saved evaluator reports and score files mark only their baselines resolved. Fresh evaluations of the unchanged patch pairs reproduce both losses. Sphinx's reviewed patch removes separate-parameter formatting from the baseline fix. SymPy's reviewed summation guard reads an unassigned local variable. Each reviewed patch fails its original \code{FAIL\_TO\_PASS} test and retains all tracked \code{PASS\_TO\_PASS} tests, respectively 40/40 and 25/25. Thus task resolution can be lost while those previously passing tests remain successful. Each pair uses the same reconstructed image. Historical image identifiers are unavailable, preventing a byte-identical environment comparison. Sphinx also has an unrelated warning-count failure in both arms outside the required test lists. The artifact retains full logs, build recipes, and infrastructure failures for these four evaluations. Original reports remain the evidence for the other 498 tasks.

Eight gains and two losses give an exact two-sided sign-test $p=0.109375$ under equal gain and loss probabilities. Only the kit route receives extra inference, and case analysis selects changed patches among shared passes. An ordinary-review control for this cohort must receive the same pre-review context, budget, and tool restrictions. Some archived baseline transcripts include the resumed review, so that context boundary must be reconstructed from the conversation. A later local run uses changed generations and remains separate from the original cohort.

A separate historical Claude study supplies an ordinary-review comparison on 45 selected SWE-bench Verified tasks. It records Opus~4.8 at max effort, read/write tools, and 200-turn caps for each arm. Both review arms inherit the same baseline messages and patch. Among 44 completed triples, the baseline and kit review each resolve 33 tasks, and ordinary review resolves 32. The single discordance gives an exact two-sided paired $p=1.0$. Kit review averages 828 seconds, 26 turns, and 62,529 output tokens. Ordinary review averages 597 seconds, 14.9 turns, and 45,898 output tokens. Equal caps therefore permit unequal consumed compute. This selected cohort provides no persuasive evidence of a kit advantage.

The tasks came from an earlier exploratory kit corpus. One baseline violates the tool restriction, leaving both review arms unrun. One completed kit arm records two attempts. Restricting to the 43 triples with one attempt per arm gives 32 kit and 31 ordinary-review successes. Kit \code{d0d07ba} accounts for 35 completed triples with 28/28/27 baseline/kit/control solves. Kit \code{cbce1cc} accounts for nine with 5/5/5. The artifact preserves all 132 evaluator reports and paired patches, missingness, and retry records. Local transcript checks find identical inherited messages between review forks and no recorded forbidden tool calls in completed reviews. Raw conversations are omitted, with source hashes and audit commitments retained. Historical image identities and earlier retry costs are unavailable. This cohort remains separate from the 500-task comparison and its missing control.

\section{Development lessons in the current implementation}
\label{app:lessons}

\begin{table}[ht]
\centering\scriptsize
\begin{tabular}{@{}>{\raggedright\arraybackslash}p{.17\linewidth}>{\raggedright\arraybackslash}p{.25\linewidth}>{\raggedright\arraybackslash}p{.25\linewidth}>{\raggedright\arraybackslash}p{.23\linewidth}@{}}
\toprule
Study & Agent/model & Resources & Endpoint and exposure\\
\midrule
HumanEval historical & Codex 0.144.6, GPT-5.6 Sol, xhigh & Per-task kit/semantics hashes, K 7.1.293, Lean 4.22.0 & Stage 2 comparison, final Stage 6 separately. Development tasks, selected repairs.\\
HumanEval audit challenge & Codex 0.155.0-alpha.16.3, GPT-5.6 Sol, high & K 7.1.337, 24 isolated packages & Author-reviewed scopes and seeds. 24 original attempts; two capacity retries.\\
KleverBench Luna controls & Codex 0.149.1, GPT-5.6 Luna, high & Fixed \code{fe49fb5} core, K 7.1.337 & 31 programs, three arms, one original attempt per cell. Four setup recoveries are separate.\\
KleverBench DeepSeek controls & Codex 0.149.1, DeepSeek-V4.1-Flash, high & Same task instructions, core and K image & 31 programs, three arms, one original attempt per cell. Provider transport and \$0.10 cash allocation differ.\\
KleverBench historical & Codex 0.149.1, GPT-5.6 Luna & Kit \code{fe49fb5}, K 7.1.337 & Four attempts, structural/coverage/proof checks. Exposure boundary unknown.\\
KleverBench extension & Same recorded Luna agent & Same kit, changed data/vocabulary & Separate nine-program cohort.\\
KleverBench DSV4 & OpenCode 1.18.23, DSV4-flash & Kit \code{fe49fb5}, K 7.1.337 & One attempt, report-derived mixed results.\\
Historical contracts & GPT-5.6 Sol; recovered records also include DeepSeek & Per-package runtime and EVM scope & Selected claims and audits; 12 final specs match recovered candidates.\\
Optimism pause & GPT-5.6 Sol, xhigh; Astra reference judge & Kit \code{8a2f727}, EVM \code{4f4c384}, London, infinite gas & Initial run plus two continuations. Scoped backend proofs and author-provided human confirmation.\\
SWE-bench & Codex 0.140.0, GPT-5.5, xhigh & Two historical kit revisions & Extra review plus official tests, no ordinary-review control.\\
SWE-bench control cohort & Claude Opus 4.8, max & Kits \code{d0d07ba} and \code{cbce1cc} & 45 selected tasks, 44 completed triples. Equal turn caps, unequal consumed compute.\\
\bottomrule
\end{tabular}
\caption{Configurations differ across studies. Recorded model aliases leave server-side snapshots unspecified. The supplementary registry supplies budgets, selection limits, and source links. Dated repairs and replays have separate provenance.}
\label{tab:configurations}
\end{table}

\begin{table}[ht]
\centering\small
\begin{tabular}{@{}lrrrl@{}}
\toprule
Kit snapshot & Skills & Bytes & Tokens & Distinct membership\\
\midrule
Historical \code{fe49fb5} & 8 & 111,020 & 24,966 & Includes on-paper reasoning\\
Released \code{5de7a09} & 8 & 118,181 & 26,202 & Includes client setup\\
Development \code{8a2f727} & 9 & 125,119 & 27,809 & Includes both\\
\bottomrule
\end{tabular}
\caption{Static sizes of the delivered Markdown skills and references. Tokens sum separately encoded files under \code{tiktoken} 0.14.0 and \code{o200k\_base}, including publication redactions and excluding message framing. Shared references add no skills. These counts measure library size. Historical Luna loading records are unavailable.}
\label{tab:kit-inventory}
\end{table}

The development entry skill handles construction and specification review in the ongoing session. It delegates final proof validation to a fresh agent when available and permitted. Otherwise, it records a same-agent review of saved artifacts under the same stage contracts. Earlier releases organize these stages differently. The supplement versions the development kit, released plugin, and historical treatment instructions separately. The September 24 kit revision adds checked summary adequacy and pinned library-model comparisons. Parent and changed-file hashes define the update, and a helper reconstructs the kit from the preserved snapshot. This revision awaits benchmark evaluation.

\begin{table}[h]
\centering\small
\begin{tabular}{@{}>{\raggedright\arraybackslash}p{.25\linewidth}>{\raggedright\arraybackslash}p{.33\linewidth}>{\raggedright\arraybackslash}p{.34\linewidth}@{}}
\toprule
Observed difficulty & Retained procedure & Evidence requested\\
\midrule
Symbolic state differs from invariant & Compare computation, bindings, and framed cells before revising mathematics & Residual state and matching repaired invariant\\
\addlinespace
Helper rule replaces source execution & Classify the extension by its effect and establish the connection & Bridge-free execution theorem over the full matched context\\
\addlinespace
Precondition removes difficult inputs & Review the complete intended domain and test coverage inputs & Rejected exclusions and documented case splits\\
\addlinespace
Storage addresses alias & Apply writes in execution order and include the same-slot case & Claims covering aliased and distinct locations\\
\addlinespace
Exception occurs after a log & State the execution boundary and model rollback at the call level & Raw-state and call-level claims\\
\addlinespace
Accepted remote job loses its connection & Retain session and task identifiers and recover evidence & Task status, cancellation, and downloaded logs\\
\bottomrule
\end{tabular}
\caption{Examples of procedures retained from the development record. The mapping documents the content of the current kit, with no isolated causal estimate for an individual procedure.}
\label{tab:lessons}
\end{table}

The examples record successful runs and workflow deviations. These include a skipped specification audit, an audit completed after proof construction, and an early audit performed by the producer after a storage incident. The resulting changes clarify artifact requirements and stage dispatch. They also identify tests for interrupted requests, slow compilation, unavailable storage, and audit-stage enforcement. We retain these as development examples outside the measured benchmark cohorts.

\end{document}